# Fast algorithms for simulating chiral fermions in U(1) lattice gauge theory

**Dafina. Xhako**[*], **Artan. Boriçi**

Department of Physics, Faculty of Natural Sciences, University of Tirana, Tirana, Albania

**Email address:**
dafina.xhako@fshn.edu.al (D. Xhako), borici@fshn.edu.al (A. Boriçi)

**Abstract:** Lattice QCD with chiral fermions are extremely computationally expensive, but on the other hand provides an accurate tool for studying the physics of strong interactions. Since the truncated overlap variant of domain wall fermions are equivalent to overlap fermions in four dimensions at any lattice spacing, in this paper we have used domain wall fermions for our simulations. The physical information of lattice QCD theory is contained in quark propagators. In practice computing quark propagator in lattice is an inversion problem of the Dirac operator matrix representing this quarks. In order to develop fast inversion algorithms we have used overlap solvers in two dimensions. Lattice QED theory with U(1) group symmetry in two dimensional space-times dimensions has always been a testing ground for algorithms. By the other side, motivated by our previews work that the two-grid algorithm converge faster than the standard iterative methods for overlap inversion but not for all quark masses, we thought to test this idea in less dimensions such as U(1) gauge theory. Our main objective of this paper it is to implement and develop the idea of a two level algorithm in a new algorithm coded in QCDLAB. This implementation is presented in the preconditioned GMRESR algorithm, as our new contribution in QCDLAB package. The preconditioned part of our algorithm, different from the one of [18], is the approximation of the overlap operator with the truncated overlap operator with finite N3 dimension. We have tested it for 100 statistically independent configurations on 32 x 32 lattice background U(1) field at coupling constant β=1 and for different bare quark masses mq = [0.5, 0.45, 0.4, 0.35, 0.3, 0.25, 0.2, 0.15, 0.1]. We have compared the convergence history of the preconditioned GMRESR residual norm with another overlap inverter of QCDLAB as an optimal one, such as SHUMR. We have shown that our algorithm converges faster than SHUMR for different quark masses. Also, we have demonstrated that it saves more time for light quarks compared to SHUMR algorithm. Our algorithm is approximately independent from the quark mass. This is a key result in simulations with chiral fermions in lattice theories. By the other side, if we compare the results of [18] for quark mass 0.1 in SU(3), results that our chosen preconditioned saves a factor of 2 but in U(1). Our next step is to test this algorithm in SU(3) and to adopt it in parallel.

## 1. Introduction

Quantum Chromodynamics (QCD) is the quantum theory of interacting quarks and gluons. It should explain the physics of strong force from low to high energies. Due to asymptotic freedom of quarks at high energies [1], it is possible to carry out perturbative calculations in QCD and thus succeeding in explaining a range of phenomena. At low energies quarks are confined within hadrons [1] and the coupling between them is strong. This requires non-perturbative calculations. The direct approach is known to be the lattice approach. The lattice regularization of gauge theories was proposed by [Wilson 1974]. It defines the theory in an Euclidean 4-dimensional finite and regular lattice with periodic boundary conditions.

Lattice regularization of chiral fermions is an important development of the theory of elementary particles. After many years of research in lattice QCD, it was possible to formulate QCD with chiral fermions on the lattice. There are two chiral formulations: a) domain wall fermions [2, 3] and b) overlap fermions [4, 5], which are closely related [6]. In particular, the truncated overlap variant of domain wall fermions [7] can be shown to be equivalent to overlap fermions in four dimensions at any lattice spacing [8].

The physical information of these theories is contained in quark propagators, which are then combined to construct meson, nucleon and other elementary particle propagators. One of the basic major computing problem in lattice QCD simulations is the calculation of quark propagator. Generally, this problem leads to very intensive computations and requires high-end computing platforms. For this purpose, simulations of lattice theory with U(1) group symmetry (a unitary symmetry of one dimension that corresponds to a symmetry under transformations of phase), or the case of the Quantum Electrodynamics, in two dimensional space-times has always been a testing ground for algorithms.

In this work we develop in fact prototypes of algorithms that are using high-end computing as our contribution to QCDLAB package [9, 10] for simulating chiral fermions. This package, a MATLAB/OCTAVE based environment, allows fast prototyping of linear algebraic computations and thus accelerates the process of finding the most efficient fermion algorithm. In practice computing quark propagator in lattice is an inversion problem of the Dirac operator matrix representing these quarks. Motivated by the results of the two grid algorithm that we proposed [11, 12] as the fastest overlap solver in SU(3) gauge theory, which didn't converges for all quark masses,  we thought to test this idea in less dimensions such as U(1) gauge theory. In our case we want to calculate the domain wall fermion propagator, but in order to develop fast algorithms, we use the truncated overlap variant of domain wall fermions in 2+1 dimensions with the extra finite dimension N3.

## 2. The Chiral Overlap Operator

Quark propagator computations amount to solving large linear systems of the type:

$$Dx = b \quad (1)$$

where $D \in \mathbb{C}^{N \times N}$ is a sparse and large matrix operator representing the Dirac operator on a regular four dimensional space-time lattice, $x, b \in \mathbb{C}^N$ are the quark propagator and it's source. In order to simulate chiral fermions on lattice QCD one can use as Dirac operator the chiral Dirac operator that is representing by the Neuberger operator, which is a shifted unitary matrix of the form [13]

$$D^N = c_1 I - c_2 V, \quad (2)$$

where $V = A(A^+A)^{-1/2}$ is a unitary matrix, $I$ unit matrice and $A = M - aD^W$. The overlap operator, $D^N$ it's non-hermitian and it is a *dense* and *large* matrice. This operator can be expressed in the equivalent way from signum function [13],

$$D^N = c_1 I - c_2 \gamma_5 sign(H_W) \quad (3)$$

where $H_W = \gamma_5(M - aD^W)$. For the signum function to be nontrivial, the Wilson-Dirac operator should be indefinite, which is the case if its bare mass $M$ is sufficiently negative and it is usually taken to be in the interval (-2,0). $a$ is the lattice parameter, $c_1$ and $c_2$ are two constants,

$$c_1 = (1+m_q)/2, \quad c_2 = (1-m_q)/2 \quad (4)$$

$m_q$ is bare quark mass and $D^W$ the Wilson-Dirac operator,

$$D^W = 1/2 \sum_\mu \left[ \gamma_\mu \left( \partial_\mu^* + \partial_\mu \right) - a \partial_\mu^* \partial_\mu \right] \quad (5)$$

with $\partial_\mu$ and $\partial_\mu^*$ as the the nearest-neighbor forward and backward difference operators. These operators are unitary 3 by 3 matrices with determinant one that are associated with links of the lattice and are oriented positively. A set of such matrices forms a "configuration". $\gamma_\mu, \mu = 1,...,5$ are 4 by 4 matrices related to the spin of the particle. Therefore, if there are *N* lattice points, the matrix is of order *12N*. In the case of U(1) lattice gauge theory the complexity of this matrix it is *2N*.

## 3. Fast Inverting Algorithms

In this paper we use a special package software named QCDLAB [9, 10] which is a design and research tool for lattice QCD algorithms. It is a collection of MATLAB functions,that is based on a "small-code" and a "minutes-run-time" algorithmic design philosophy.  In this work we use the QCDLAB 1.0 version [9] with the Schwinger model on the lattice, a great simplification, which shares many features and algorithms with lattice QCD. In QCDLAB 1.0, the overlap operator $D^N$ can be computed from the inverse square root of $A^*A$:

$$\left(A^*A\right)^{-1/2} = V\Sigma^{-1}V^* \quad (6)$$

where $\Sigma$ are the singular values of $A = U\Sigma V^*$. Since we require only the quark propagator of overlap chiral fermions, the full matrix $D^N$ is not required. The multiplication of $D^N$ with a vector can be computed using Krylov subspace algorithms [14] such as the double pass Lanczos algorithm [15] and Zolotarev approximation [16] for the inverse square root of the Lanczos matrix *T*. The algorithm requires a lower bound *lam*bda of the smallest eigenvalue of $A^*A$. This idea is implemented in SHUMR (Shifted Unitary Minimal Residual) optimal algorithm which has been part of QCDLAB 1.0., and it is shown in the Appendix C. This algorithm is tested before also in case of SU(3) gauge theory and gives good results [17]. Based on the idea of a two level algorithm as proposed in [11, 12], in this work we develop a faster algorithm, the preconditioned GMRESR [18] (Generalized Minimal Residual Method - Recursive) algorithm developed   it as part of QCDLAB 1.0 package, in U(1) gauge field background. So in this way we have added in QCDLAB 1.0 new efficient routines.

In reference [18] the preconditioner is built upon an inaccurate approximation to the sign function, while we use as the preconditioned part the approximation of the overlap operator with the truncated overlap operator with finite N3 dimension. The preconditioned GMRESR algorithm uses the truncated overlap operator along $N_3 = 8$, $D^{TOV}$, not the real overlap operator as SHUMR, so maybe we lose the accuracy of the exact overlap operator but we gain from the low complexity of the truncated overlap version. In order to control the residual norm of this algorithm we use the true residual norm from the exact solution. In general for the solution of the linear system $Dx = b$ we investigate this condition in the $k$-th step,

$$\underbrace{\left\| b - Dx^k \right\|}_{\substack{true \\ residual}} \leq \underbrace{\left\| r^k - \left( b - Dx^k \right) \right\|}_{\substack{residual \\ gap}} + \underbrace{\left\| r^k \right\|}_{\substack{computed \\ residual}} \quad (7)$$

and develop a strategy to bound residual gap below required accuracy. If the number of iterations to reach the desired residual reduction is large, then there can be a considerable accumulation of the errors in the matrix-vector product in the residual gap. In practice, this might mean that the tolerance on the matrix-vectors has to be decreased. In essence the preconditioned GMRESR follows two levels;
*First level:* The preconditioned part: Compute approximate solution for the linear system using the equivalent of the overlap operator, the truncated overlap operator in 2 + 1 dimensions with an extra finite $N_3$ dimension using small accuracy *tol0*. In this case we have used the CGNE (Conjugate Gradients on Normal Equations) algorithm as an known optimal solver [19].

*Second level:* Find the true residual using the real overlap operator in $N_3$ infinite dimension, and use this residual to control the residual of the first step in each iteration. The Matlab/Octave code function for this second step we called Mult_Overlap.m and it is shown in the Appendix B. This cycle will continue still is reached a desired tolerance *tol*.

The Matlab/Octave code function of the preconditioned GMRESR algorithm is shown in the Appendix A. Below we are giving briefly the preconditioned GMRESR algorithm for inverting truncated overlap operator in 2+1 dimension, so in U(1) gauge theory.

**ALGORITHM: The preconditioned GMRESR (A, $D^{TOV}$, b, $c_1$, $c_2$, tol)**

#computes $x$ with $\left\| Dx - b \right\| \leq tol \cdot \left\| b \right\|$

# c1, c2 the coefficients of the overlap operator

# inital values

$x = 0$;

$r = b$;

#empty matrix

$C = [\,]$;

$U = [\,]$;

#outer iteration, the preconditioned part

**while** $\left\| r \right\| > tol \cdot \left\| b \right\|$ **do**

**solve** $D^{TOV} u = r$, *to relative accuracy* $tol0$,

**ALGORITHM: The preconditioned GMRESR (A, $D^{TOV}$, b, $c_1$, $c_2$, tol)**

*from* $u = cgne(D^{TOV}, r, tol0)$;

#inner iteration, compute the true residual from c with

# $\left\| D^N u - c \right\| \leq tol \cdot \left\| b \right\| \cdot \left\| u \right\| / \left\| r \right\|$

$c = c_1 \cdot u + c_2 \cdot (V \cdot u)$, where $V \cdot u$ is computed from

$V \cdot u = Mult\_Overlap(A, u)$;

**for** $i = 1 : (C, 2)$ **do**

$\alpha = C[:,i]^{\dagger} \cdot c$;

$c = c - \alpha \cdot C[:,i]$;

$u = u - \alpha \cdot U[:,i]$;

**end for**

# calculate

$c = c / \left\| c \right\|$; $u = u / \left\| c \right\|$;

$C = [C, c]$; $U = [U, u]$;

$\gamma = c^{\dagger} \cdot r$; $x = x + \gamma \cdot u$; $r = r - \gamma \cdot c$;

**end while**

## 4. Results and Discussion

We have made simulation for 100 statistically independent configurations on 32 x 32 lattice background U(1) field at coupling constant $\beta = 1$ and for different bare quark masses $m_q$ = [0.5, 0.45, 0.4, 0.35, 0.3, 0.25, 0.2, 0.15, 0.1]. The third dimension for truncated overlap fermions we set is $N_3 = 8$ and $tol0 = 0.1$; $tol = 10^{-6}$. The range parameter $M$ we have fixed at -0.345 and we have taken the lower bound *lambda* = $10^{-7}$ which requires the Zolotarev rational polynomial to be of the order $n = 60$, i.e. the number of $T$ inversions is thus $n/2 = 30$. We compare the convergence history of the preconditioned GMRESR residual norm with SHUMR. The graphical results of the history of the convergence for each algorithm shows that our algorithm converges faster than SHUMR for different quark masses (Fig.1, Fig.2, Fig.3). As we can see from Fig.1, for quark mass 0.5 our algorithm saves a factor of 2, from Fig.2 for quark mass 0.3 it saves a factor of 3 and from Fig.3 for quark mass 0.1 it saves a factor of 7. So, what is more important is the fact that our algorithm is faster and we gain more time for light quarks. Also, the Fig.3 shows that our algorithm in two-dimension saves a factor of 2 compare to the results of [18] for the same quark mass but in SU(3) gauge theory. This result must be checked for our algorithm in four dimension in our future work. We have calculated also the inverting time required till the convergence for each algorithm for different quark masses that is shown in Fig. 4.

## 5. Conclusions

The chiral symmetry in lattice QCD is crucial because this property is fundamental for the strong interactions. Simulations in lattice with chiral fermions have high computation cost because of the complex form of operator that is related to them, the overlap operator. We showed that our algorithm, the preconditioned GMRESR, converges faster compared to an optimal solver of overlap operator, such as SHUMR, for different quark masses. Also, we showed that it saves much more time for light quarks compared to other algorithm and that it is approximately independent from the quark mass. In this way our algorithm avoid what it is called critical slowing down of the algorithms with light quarks. This is a key result in simulations with chiral fermions in lattice theories. From the results obtained seems that our algorithm is very promising and in our further studies we want to adopt this algorithm in parallel.

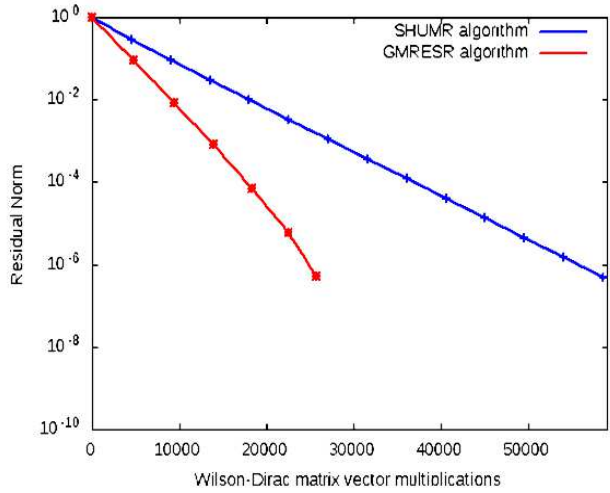

*Figure 1.* The history of the convergence of the norm of residual as the function of the number of $D_W$ multiplications for SHUMR and the preconditioned GMRESR inverting overlap operator on 32 x 32 lattice background U(1) field at coupling constant $\beta = 1$ and quark mass 0.5(in lattice unit).

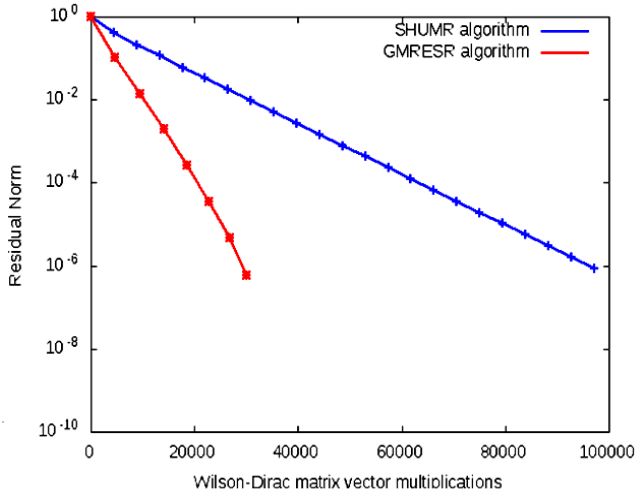

*Figure 2.* The history of the convergence of the norm of residual as the function of the number of DW multiplications for SHUMR and the preconditioned GMRESR inverting overlap operator on 32 x 32 lattice background U(1) field at coupling constant $\beta = 1$ and quark mass 0.3(in lattice unit).

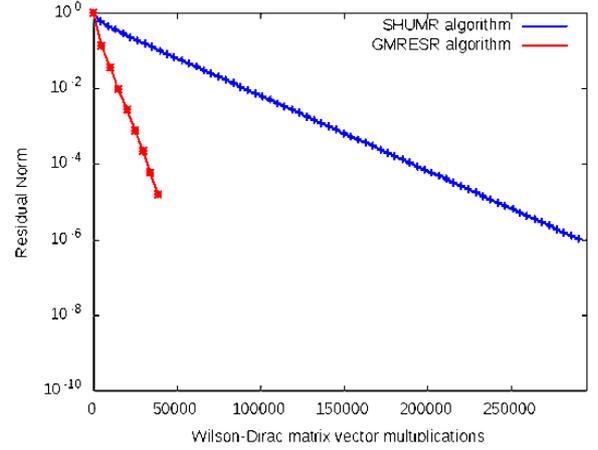

*Figure 3.* The history of the convergence of the norm of residual as the function of the number of DW multiplications for SHUMR and the preconditioned GMRESR inverting overlap operator on 32 x 32 lattice background U(1) field at coupling constant $\beta = 1$ and quark mass 0.1(in lattice unit).

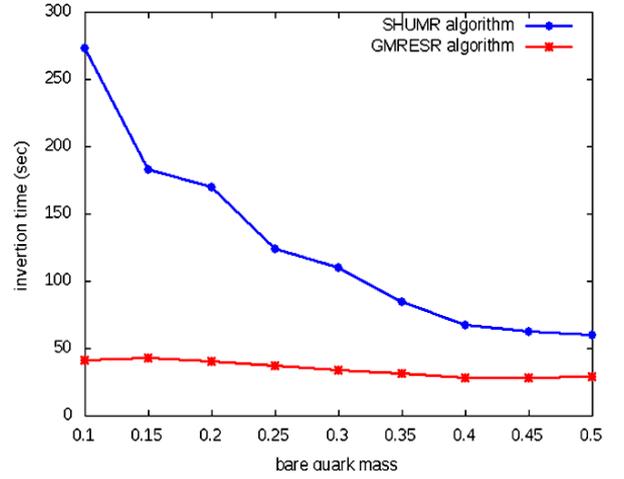

*Figure 4.* The inverting time of the overlap operator as function of the quark mass (in lattice unit) for SHUMR and the preconditioned GMRESR algorithms on 32 x 32 lattice background U(1) field at coupling constant $\beta = 1$.

## Appendix A

The preconditioned GMRESR code function for inverting truncated overlap operator in U(1) gauge theory.

```
function[x,Matvec,rr]=gmresr(A,psi,lambda,c1,c2,
M,M1,P,b,tol0,tol,nmax)
n=max(size(b));    N3=max(size(P))/n;    matvec=0;
Matvec=matvec;
k=1; b=b(:); r=b; x=zeros(n,1);
rnorm=norm(r); rr=rnorm; U=[]; C=[];
while ((rnorm>tol)&&(k<=nmax));
% the preconditioned part, compute approximate solution using the approximate overlap operator the so called truncated overlap oeprator with accuracy tol0
    r1=zeros(n*N3,1); r1(1:n)=r;
    r1=M1*(P*r1);
```

```
    [u1,rrM]=cgne(M,r1,r1,tol0*rnorm,n*N3);
    multDw=2*N3*max(size(rrM));
    u=P'*u1; u=u(1:n);
% Find the true residual using the exact overlap operator
    [Vu,multA]=Mult_overlap(A,u,psi,lambda,n*N3);
    c=c1*u+c2*Vu;
    for i=1:k-1;
        alpha=C(:,i)'*c; c=c-alpha*C(:,i); u=u-alpha*U(:,i);
    end
    beta=norm(c);
    c=c/beta; u=u/beta; U=[U,u]; C=[C,c];
    gamma=c'*r;
    x=x+u*gamma;
    r=r-c*gamma;
    rnorm=norm(r); rr=[rr;rnorm];
    matvec=matvec+multDw+multA;
Matvec=[Matvec,matvec];
    k=k+1;
end
```

## Appendix B

Mult_Overlap code function using exact overlap operator

```
function
[Vx,Matvec,rr]=Mult_overlap(A,x,psi,lambda,nmax)
% Given the Dirac operator A (shifted adequately) this function
% multiplies x with the unitary part U*V' of A=U*S*V'
% using Lanczos algorithm to compute inverse square root of A'*A
% and Zolotarev approximation for the inverse suqare root of T
% uses proejction of smallest eigenpairs (psi,lambda) of A'*A
Matvec=0;
AHA=A'*A;
Qx=x-psi*(psi'*x); % projected right hand side
[T1,matvec,rr]=lanczos(AHA,Qx,psi,lambda,1e-6,nmax,[],[],0);
Matvec=Matvec+matvec;
T=T1;T(end,:)=[];
% compute 1/sqrt(T)*e1
epsilon=max(real(diag(lambda))); % smallest eigenvalue of A'*A
n=60; % order of Zolotarev approximation
[d,q,y,Delta]=coef_zolotarev(1,n,epsilon);
nlanczos=max(size(T));
e1=zeros(nlanczos,1);e1(1)=1;
d0=d(1);d(1)=[];n2=max(size(d));
inv_sqrt_T=d0*e1;
for l=1:n2;
    Tl=sparse(epsilon*speye(nlanczos)+T*q(l));
    inv_sqrt_T=inv_sqrt_T+T*d(l)*(Tl\e1); % n2 inversions of Tl
end
inv_sqrt_T=inv_sqrt_T*norm(x);
% second call to lanczos
[y,matvec]=lanczos(AHA,Qx,psi,lambda,1e-6,nmax,T1,inv_sqrt_T,1);
Matvec=Matvec+matvec;
inv_sqrt_lambda=diag(1./sqrt(real(diag(lambda))));
y1=y/sqrt(epsilon)+psi*(inv_sqrt_lambda*(psi'*x));
Vx=A*y1;
```

## Appendix C

The SHUMR code function for inverting overlap operator in U(1) gauge theory.

```
function[x,Matvec,rr]=SHUMR(A,psi,lambda,b,c1,c2,tol,imax);
b=b(:); N=max(size(b)); vzero=zeros(N,1); x=vzero;
r=b; matvec=0; Matvec=matvec;
rho = norm(r); rnorm=rho; rr=rho; alpha = rho;
u12=0; beta=1; L11_tilde=1; q=r/rho;
q_old=vzero; v_old=vzero; w_old=vzero; s_old=vzero;
% start added lines
c_km1=1; s_km1=0; c_km2=0; s_km2=0;
p1=vzero; p2=vzero; Dp1=vzero; Dp2=vzero;
% end added lines
counter = 1;
while ( (rnorm > tol) & (counter<=imax) );
    [v,multDw]=Mult_overlap(A,q,psi,lambda,imax);
    if (counter > 1),
        u12=-(q_old'*v)/(q_old'*v_old);
    end
    gamma=-c1*u12;
    L11=(q'*v)+u12*(q'*v_old);
    q_tilde=v-L11*q+u12*v_old;
    L21=norm(q_tilde);
    if (L21<=tol), break, end;
    w=q+q_old*u12+w_old*gamma/L11_tilde;
    s=c1*(q+q_old*u12)+c2*(v+v_old*u12)+s_old*gamma/L11_tilde;
    L11_tilde=c1+c2*L11-beta*gamma/L11_tilde;
    alpha=alpha*beta/L11_tilde;
    x=x+w*alpha;
    r=r-s*alpha;
    q_old=q;
    v_old=v;
    w_old=w;
    s_old=s;
    q=q_tilde/L21;
    beta=-c2*L21;
    rnorm=norm(r);
% start added lines
    t11=c1+c2*L11;
    mu=t11*c_km1+gamma*conj(s_km1)*c_km2;
    nu=c2*L21;
    if (mu != 0),
```

```
    c_k=abs(mu)/sqrt(abs(mu)*abs(mu)+abs(nu)*abs(nu));
    s_k=conj(c_k*nu/mu);
    else
    c_k=0;
    s_k=1;
    end
    omega=nu*alpha*s_k;
    mu_k=c_k*mu+s_k*nu;
    eps=t11*s_km1-gamma*c_km1*c_km2;
    theta=-gamma*s_km2;
    p=(q+q_old*u12-p1*eps-p2*theta)/mu_k;
Dp=(c1*(q+q_old*u12)+c2*(v+v_old*u12)-Dp1*eps-Dp2
*theta)/mu_k;
    rnorm_p=norm(r+omega*Dp);
    xp=x-omega*p;
    c_km2=c_km1; s_km2=s_km1; p2=p1; Dp2=Dp1;
    c_km1=c_k; s_km1=s_k; p1=p; Dp1=Dp;
% end added lines
    rr=[rr;rnorm];
    matvec=matvec+multDw;
    Matvec=[Matvec,matvec];
    counter++;
end
```

~~~~~~~~~~~~~~~~~~~~~~~~~~~~~~~~~~~~~~~~~~~~~~~~~